\documentclass[preprint]{aastex}

\begin{document}

\title {The Varying Light Curve and Timings of the Ultra-short Period Contact Binary KIC 9532219 }
\author{Jae Woo Lee$^{1,2}$, Kyeongsoo Hong$^1$, Jae-Rim Koo$^1$, and Jang-Ho Park$^{1,3}$ }
\affil{$^1$Korea Astronomy and Space Science Institute, Daejeon 34055, Korea}
\affil{$^2$Astronomy and Space Science Major, Korea University of Science and Technology, Daejeon 34113, Korea}
\affil{$^3$Department of Astronomy and Space Science, Chungbuk National University, Cheongju 28644, Korea}
\email{jwlee@kasi.re.kr, kshong@kasi.re.kr, koojr@kasi.re.kr, pooh107162@kasi.re.kr}

\begin{abstract}
KIC 9532219 is a W UMa-type eclipsing binary with an orbital period of 0.1981549 d that is below the short-period limit 
($\sim$0.22 d) of the period distribution for contact binaries. The {\it Kepler} light curve of the system exhibits 
striking light changes in both eclipse depths and light maxima. Applying third-body and spot effects, the light-curve synthesis 
indicates that the eclipsing pair is currently in a marginal contact stage with a mass ratio of $q$=1.20, an orbital inclination 
of $i$=66.0 deg, a temperature difference of $\Delta$ ($T_{1}$--$T_{2}$)=172 K, and a third light of $l_3$=75.9 \%. To understand 
the light variations with time, we divided up the light curve into 312 segments and separately analyzed them. The results reveal 
that variation of eclipse depth is primarily caused by changing amounts of contamination due to the nearby star KIC9532228 
between the {\it Kepler} Quarters and that the variable O'Connell effect originates from the starspot activity on 
the less massive primary component. Based on our light-curve timings, a period study of KIC 9532219 indicates that 
the orbital period has varied as a combination of a downward parabola and a light-travel-time (LTT) effect due to a third body, 
with a period of 1196 d and a minimum mass of 0.0892 $M_\odot$ in an eccentric orbit of 0.150. The parabolic variation could be 
a small part of a second LTT orbit due to a fourth component in a wider orbit, instead of either mass transfer or 
angular momentum loss. 
\end{abstract}

{}

\section{INTRODUCTION}

W UMa-type contact binaries consist of two dwarf stars surrounded by a common convective envelope and are recognized by 
continuous brightness variations over time and nearly equal minima in the light curve. They are thought to have evolved from 
initially detached binaries by the combined action of magnetic braking and tidal friction, and to end with coalescence of 
both components into single stars (Bradstreet \& Guinan 1994; Eggleton \& Kiseleva-Eggleton 2002). This process could only 
happen if they were very close binaries to start with and the orbital angular momentum was tidally coupled to the spin angular 
momentum. In this scenario, a circumbinary component may have played an important role in the formation of 
the initial tidal-locked close binaries with an orbital period of several days, through energy and angular momentum exchanges 
(Eggleton \& Kisseleva-Eggleton 2006; Fabrycky \& Tremaine 2007). Tokovinin et al. (2006) found that 96 \% of the close binaries 
with periods shorter than 3 d exist in multiple systems. The statistical study by Pribulla \& Rucinski (2006) indicated that 
a large proportion of W UMa binaries have circumbinary companions. The results suggest that circumbinary objects are necessary 
for the formation and evolution of short-period close binaries.

It is known that contact binaries have a very short period cut-off at about 0.215$-$0.22 d but there is no full explanation 
for this limit (Rucinski 1992, 2007). Binary stars with orbital periods shorter than 0.22 d are extremely rare 
(e.g. Dimitrov \& Kjurkchieva 2015) and most of them appear to be triple star systems (Jiang et al. 2015; Qian et al. 2015). 
Such short-period binaries provide significant information about the origin and evolutionary processes of the systems. 
The presence of a third body around an eclipsing binary could be detected by detailed analyses of both light curves and 
eclipse timings. To advance this subject, we have been studying W UMa-type eclipsing binaries below the period limit. 

KIC 9532219 (RA$_{2000}$=19$^{\rm h}$36$^{\rm m}$52$\fs018$; DEC$_{2000}$=+46$^{\circ}$10${\rm '}$48$\farcs$80; $K_{\rm p}$=$+$16.118; 
$g$=$+$16.861; $g-i$=$+$1.061) with a period of 0.198155 d is one of the shortest period W UMa-type binaries currently known. 
From an analysis of initial {\it Kepler} observations, Kjurkchieva \& Dimmitrov (2015) reported that the variable star is not 
a $\delta$ Sct pulsator but a contact binary with a mass ratio of $q$=0.9320, an orbital inclination of $i$=68.42 deg, 
a temperature difference of $\Delta T$=83 K between the components, and a third light of $l_3$=78 \%. The orbital period of 
the system was examined by Conroy et al. (2014) using the {\it Kepler} times of minimum lights between Quarters 3 and 16. 
They suggested that the cause of the period change is a light-travel-time (LTT) effect due to a third body with a period of 
$P_3$=1062.1$\pm$76.3 d and an eccentricity of $e_3$=0.372$\pm$0.002. In this paper, we present and discuss the variations in 
the light curves and eclipse timings of KIC 9532219 below the short-period limit using the {\it Kepler} data.

\section{{\it KEPLER} PHOTOMETRY AND LIGHT VARIATIONS}

The {\it Kepler} observations of KIC 9532219 were obtained during Quarters 3$-$17 in long cadence (LC) mode with a sampling rate 
of 29.4 min. We downloaded and analyzed the original data from the {\it Kepler} archive, but the results are not much different 
from the detrended light curves in the {\it Kepler} EB catalogue (Pr\v sa et al. 2011; Slawson et al. 2011)\footnote{http://keplerebs.villanova.edu/}.
Further, because the PDC detrending is optimized for planet transits, its effects on binary stars are known to be adverse in 
a significant fraction of all cases (Pr\v sa et al. 2011). Thus, we used the data in the EB Catalogue detrended from the raw SAP 
time series. The light curve is plotted in Figure 1 as magnitude versus BJD. As shown in the figure, the binary star appears to 
display significant changes in the depths of the eclipses. Because the sample rate of the {\it Kepler} data corresponds to 
about 10\% of the binary orbital period, it is impossible to measure a reliable magnitude at a given phase for each light curve. 
Thus, to look for any light variations of KIC 9532219, we combined the {\it Kepler} data at intervals of 20 orbital periods 
(20$\times$$P$=3.9631 d) and made a total of 312 light curves. Then, we measured the light levels at four characteristic phases 
(Min I at phase 0.0, Max I at phase 0.25, Min II at phase 0.5, and Max II at phase 0.75) for the separate datasets. These are 
listed in Table 1, where the mean time in the first column was computed by averaging starting and ending BJDs of the observations. 
Their errors are the standard deviations relative to the mean value of each dataset.  

In Figure 2, the magnitude differences among the measurements are displayed as Max I $-$ Min I, Max I $-$ Min II, 
Max II $-$ Min I, Max II $-$ Min II, Max I $-$ Max II, and Min I $-$ Min II. As seen in this figure, the eclipse depths, 
given in the first through fourth panels, varied significantly with time in almost identical patterns. The values of 
Max I $-$ Max II indicate that the light curves around approximately BJD 2455200 present equal light levels at the quadratures, 
while before that, Max I is brighter than Max II, and after that, the brightness differences show the inverse O'Connell effect 
with Max I fainter than Max II. These phenomena could be produced by dynamical interactions due to circumbinary companions 
gravitationally bound to the eclipsing pair (Torres \& Stefanik 2000; Zasche \& Paschke 2012) and/or stellar activity such as 
starspots (Kang et al. 2002; Lee et al. 2010). Further, the light variations are affected by the presence of 
a neighboring star KIC 9532228 (RA$_{2000}$=19$^{\rm h}$36$^{\rm m}$52$\fs418$; DEC$_{2000}$=+46$^{\circ}$10${\rm '}$48$\farcs$82; 
$K_{\rm p}$=$+$16.928; $g$=$+$17.746; $g-i$=$+$1.169) with a separation of about 4 arcsec, because the plate scale of 
the {\it Kepler} CCD camera is 3.98 arcsec pixel$^{-1}$ (Koch et al. 2010).

\section{LIGHT-CURVE SYNTHESIS AND THE CHANGE OF ECLIPSE DEPTH}

Figure 3 shows the {\it Kepler} light curve of KIC 9532219 distributed in orbital phase instead of BJD as in Figure 1. Its shape 
is typical of a short period W UMa-type binary and the curved bottoms of both minima indicate partial eclipses. To determine 
the physical parameters of the system, the {\it Kepler} data was analyzed using the 2007 version of the Wilson-Devinney binary code 
(Wilson \& Devinney 1971, van Hamme \& Wilson 2007; hereafter W-D). For this purpose, the mean light level at phase 0.75 was set 
to unity. The light-curve synthesis was performed in a manner identical to that for the eclipsing systems V404 Lyr (Lee et al. 2014) 
and KIC 5621294 (Lee et al. 2015) exhibiting LTT effects. The effective temperature of the more massive star was set to be 5,031 K 
from the {\it Kepler} Input Catalogue (KIC; Kepler Mission Team 2009). In Table 2, parentheses signify adjusted parameters. 
To avoid possible confusion, we refer to the primary and secondary stars as those being eclipsed at Min I and Min II, respectively.  

A photometric solution of KIC 9532219 has been reported only by Kjurkchieva \& Dimmitrov (2015) and no spectroscopic observations 
have been made for the system. Thus, we used the so-called $q$-search method for various modes of the W-D code to obtain 
photometric solutions. The behavior of the weighted sum of the squared residuals, $\Sigma W(O-C)^2$, was used to estimate 
the potential reality of each model. For this procedure, 1,000 normal points were formed from the individual {\it Kepler} data, 
and weights were assigned to the number of observations per normal point. The $q$-search results indicated that both components 
are in marginal contact with respect to the inner Roche lobe. As displayed in Figure 4, the optimal solution is around $q$=1.20, 
which indicates that KIC 9532219 is a W-subtype (defined observationally by Binnendijk 1970) contact binary. 
In subsequent calculations, we treated the mass ratio ($q$) and third-body parameters ($a^{\prime}$, $e^{\prime}$, $\omega^{\prime}$, 
$P^{\prime}$, and $T_{\rm c}^{\prime}$) as adjustable variables and fitted all {\it Kepler} data simultaneously. 

The unspotted solution is listed in columns (2)--(3) of Table 2 and appears as a dashed curve in the top panel of Figure 3.
The light residuals from the model are plotted in the second panel of Figure 3, wherein it can be seen that the model 
light curves do not fit the observed ones around phase 0.25. Light asymmetry has been reported commonly for light curves 
of short-period binaries and may be due to the spot activity on stellar photospheres. Because there is currently no way 
to know which spot model is more appropriate to explain the light variation, we applied a magnetic cool spot on either of 
the component stars. In the W-D code, each spot is described by four parameters: latitude ($\theta$) and longitude ($\lambda$) 
of a spot center, and angular radius ($r_{\rm s}$) and temperature factor ($T_{\rm s}$=$T$$\rm _{spot}$/$T$$\rm _{local}$) of 
a spot. Although it is not easy to distinguish between the two spot models from only the light-curve synthesis, 
the cool spot on the primary star gives a slightly smaller value of $\Sigma W(O-C)^2$ than that on the secondary component. 
The spot solution is given in columns (4)--(5) of Table 2 together with the spot parameters. The synthetic light curve is 
plotted as the solid curve in the top panel of Figure 3 and the residuals from the spot model are plotted in the third panel  
of the figure. 

To explore in detail the light variations of KIC 9532219, including the change in the eclipse depths, we separately analyzed 
the 312 light curves combined at intervals of 20 orbital periods. For the computations, we used the spot solution as 
initial values and adjusted all the other parameters except for the orbital period ($P$) and the spot latitude ($\theta$). 
The results are listed in Table 3, and the variations in the binary and spot parameters are displayed in Figures 5 and 6, 
respectively. The light residuals from the analyses are displayed in the bottom panel of Figure 3. We can see that 
the unmodeled light variations around both eclipses have almost disappeared. As shown in Figure 5, the luminosity parameters 
have quasi-periodically changed, wherein the change of $l_{1}$ is nearly 180$^\circ$ out of phase with that of $l_{3}$ and 
displays a pattern and sense almost identical to the observed changes of the eclipse depths seen in Figure 2. On the contrary, 
the variations of the other binary parameters are relatively small and do not contribute the depth change significantly. 
In Figure 6, it can be seen that the brightness differences between the light maxima in the fifth panel of Figure 2 are 
dominated by the changes in the spot parameters with time, especially in the longitude and temperature factor of that spot. 
Even the spot variability cannot describe the change in the eclipse depth.

The observed depth of the eclipse may be affected by changing amounts of contamination due to the nearby star KIC 9532228 
because different photometric apertures may be used for different {\it Kepler} Quarters. On this account, the $l_{1}$ and 
$l_{3}$ parameters appear to have varied in complex ways rather than in monotonic fashions, unlike real values. To examine 
any periodicity for the luminosities in Table 3, we separately fitted the two parameters to a sine curve, as follows: 
\begin{equation}
 l_{s} = l_{0s} + K_{ls} \sin({{2 \pi} \over P_{ls}} t + \omega_{0s} ), ~~ s = 1 {\rm ~or ~3},
\end{equation}
where $t$ is the mean time of each luminosity. The results are plotted with the solid curves in the fifth to sixth panels of 
Figure 5, where a possible modulation seems to exist. The zero point, semi-amplitude, period, and phase of the sinusoids 
are calculated to be $l_{01}=0.1195\pm0.0004$ and $l_{03}=0.7589\pm0.0008$, $K_{l1}=0.0050\pm0.0006$ and $K_{l3}=0.010\pm0.001$, 
$P_{l1}=372\pm7$ d and $P_{l3}=368\pm7$ d, and $\omega_{01}=229\pm101$ deg and $\omega_{03}=360\pm102$ deg, respectively. 
The modulation periods of $l_{1}$ and $l_{3}$ are approximately equal to the {\it Kepler} orbit of 372.5 d. These imply that 
the observed eclipse depth and, thus, the calculated luminosities may be modulated at the {\it Kepler} orbital period due to 
the unique features of the {\it Kepler} Earth-trailing heliocentric orbit. Meanwhile, the third light ($l_3$=76\%) can come 
from both the close neighbor separated by 4 arcsec and the circumbinary companions inferred from our eclipse timing analysis 
which will be discussed in the following section. Because KIC 9532228 is about 0.81 mag fainter than KIC 9532219 in 
the {\it Kepler} passband, the nearby star may not be a dominant source of the light contribution. Then, it is possible that 
a large percentage of the third light originates from the circumbinary objects orbiting the eclipsing binary.
 
Absolute dimensions for KIC 9532219 can be estimated from our photometric solutions with the cool-spot model in Table 2 and 
the empirical relations between spectral type (temperature) and stellar mass. Assuming that the more massive secondary star is 
a normal main-sequence one with a spectral type of about K1 and, hence, a mass of $M_2$=0.86 $M_\odot$ from its temperature, 
we computed the absolute parameters of each component listed in Table 4. The luminosity ($L$) and bolometric magnitudes ($M_{\rm bol}$) 
were computed by adopting $T_{\rm eff}$$_\odot$=5,780 K and $M_{\rm bol}$$_\odot$=+4.73 for solar values. In the mass-radius, 
mass-luminosity, and HR diagrams, the locations of both components of KIC 9532219 lie near the zero-age main sequence.

\section{ECLIPSE TIMING VARIATION AND ITS IMPLICATIONS}

Times of minimum lights could be shifted from the real conjunctions due to asymmetrical eclipse minima originating from starspot 
activity (cf. Lee et al. 2015). The light-curve synthesis method developed by W-D can gives more and better information with 
respect to other methods (Maceroni \& van't Veer 1994). As an example, the method of Kwee \& van Woerden (1956; hereafter KW), 
which has been widely used for the determination of minimum epochs, does not consider spot activity and measures an eclipse time 
based on the observations during minimum alone. When the minimum is asymmetric, the uncertainty of the KW time is systematically 
increased. In the previous section, we modeled the 312 segments formed from the {\it Kepler} light curve by applying 
the time-varying cool spot on the primary component of KIC 9532219. As the result, we measured the minimum epochs for 
each dataset given in the first column of Table 3 and they were usde for ephemeris computations of the system.  

First of all, we examined whether the period change of KIC 9532219 could be represented by a single LTT ephemeris suggested 
by Conroy et al. (2014). Fitting our light-curve timings to that ephemeris form failed to give a satisfactory result. Instead, 
we found that the eclipse timing variation is best fitted by the combination of a downward-opening parabola and an LTT effect 
caused by the presence of a third body in the system as follows:
\begin{eqnarray}
C = T_0 + PE + AE^2 + \tau_3,
\end{eqnarray}
where $\tau_{3}$ is the LTT due to a circumbinary companion (Irwin 1952) and includes five parameters ( $a_{\rm b} \sin i_3$, 
$e_{\rm b}$, $\omega_{\rm b}$, $n_{\rm b}$, and $T_{\rm b}$). Here, $a_{\rm b} \sin i_3$, $e_{\rm b}$ and $\omega_{\rm b}$ 
are the orbital parameters of the eclipsing pair around the mass center of the triple system. The parameters $n_{\rm b}$ and 
$T_{\rm b}$ denote the Keplerian mean motion of the mass center of the binary system and its epoch of periastron passage, 
respectively. In this analysis, the Levenberg-Marquart technique (Press et al. 1992) was applied to solve for the eight 
unknown parameters of the ephemeris. The results are listed in Table 5 together with other related quantities. In this table,
$P_{\rm b}$ and $K_{\rm b}$ indicate the period and semi-amplitude of the LTT orbit, respectively. Within errors, 
most parameters from the eclipse timings are in good agreement with those obtained from the light curve using the W-D code.

The eclipse timing diagram of KIC 9532219 constructed with the linear terms in Table 5 is plotted in the top panel of Figure 7,
together with the measures of Conroy et al. (2014) for comparison. Here, the solid curve and the dashed parabola represent 
the full contribution and the quadratic term, respectively. The middle and bottom panels display the LTT orbit ($\tau_{3}$) and 
the residuals from the full ephemeris, respectively. As displayed in this figure, the quadratic {\it plus} LTT ephemeris currently 
provides a good representation of all eclipse times. The mass function of the third body becomes $f(M_{3})$=0.000255 $M_\odot$, 
and its mass is $M_{3} \sin i_{3}$=0.0892 $M_\odot$. If it is a main-sequence star, the minimum radius and temperature of 
this object are calculated to be $R_3$=0.0976 R$_\odot$ and $T_3$=3058 K, respectively, using the correlations from 
eclipsing binary stars (Southworth 2009). These correspond to the bolometric luminosity of $L_3 \ga$ 0.0007 L$_\odot$, which 
would contribute only $L_3 / (L_1 + L_2 + L_3 ) \ga 0.0015$ to the total light of this system. Thus, the tertiary companion 
inferred from the timing study may not be the third light source ({\it $l_{3}$}$\sim$76\%) detected in our light-curve synthesis. 
On the other hand, the eclipse timing variation may be partly caused by the perturbative effect of the circumbinary companion 
added to the pure geometrical LTT effect (Borkovits et al. 2003; \" Ozdemir et al 2003). We computed the semi-amplitude of 
the third-body dynamic perturbation on the binary orbit to be about 0.03 s, and found that its contribution to 
the timing variation is not significant.

The quadratic term ($A$) of the ephemeris represents a continuous period decrease with a rate of 
$-$5.27$\times$10$^{-7}$ d yr$^{-1}$. This corresponds to a fractional period change of $-$1.44$\times$10$^{-9}$, which agrees 
well with the value of $-$1.4$\times$10$^{-9}$ calculated with the W-D code. Such a variation can be explained by either 
mass transfer from the more massive secondary to the primary star or angular momentum loss (AML) due to a magnetic stellar wind. 
If the parabolic variation is produced by conservative mass transfer, the transfer rate is 3.92$\times$10$^{-6}$ M$_\odot$ yr$^{-1}$, 
among the largest rates for W UMa systems (cf. Lee et al. 2009). Assuming that the secondary star transfers its present mass to 
the less massive primary component on a thermal time scale $\tau_{\rm th}$ (Paczy\'nski 1971), $\tau_{\rm th}$ = 1.40$\times 10^{8}$ yr 
and mass is transferred to the companion at a rate of $M_{\rm 2}/ \tau_{\rm th}$ = 6.15$\times 10^{-9}$ M$_{\sun}$ yr$^{-1}$. 
This value is three orders of magnitude too small to be the single cause of the observed period change. 

Another possible mechanism for the parabolic variation is AML caused by magnetic braking. Guinan \& Bradstreet (1988) derived 
an approximate formula for the period decrease rate due to spin-orbit-coupled AML of binary systems, as follows:
\begin{equation}
  {dP \over dt} \approx -1.1 \times 10^{-8} q^{-1} (1+q)^2 (M_{\rm 1}+M_{\rm 2})^{-5/3} k^2 (M_{\rm 1} R^4 _{\rm 1} + M_{\rm 2} R^4 _{\rm 2} ) P^{-7/3},
\end{equation}
where $k^2$ is the gyration constant. With $k^2$=0.1 (see Webbink 1976) and with the absolute dimensions in Table 4, we computed 
the AML rate to be d$P$/d$t$ = $-$2.32 $\times 10^{-8}$ d yr$^{-1}$. This computed value is more than 22 times smaller than 
the observed one, so the AML hypothesis is not confirmed. These mean that both mass transfer and AML cannot describe the parabolic 
variation. The downward parabola may only be the observed part of a second LTT orbit ascribed to the presence of a fourth object.

\section{SUMMARY AND DISCUSSION}

In this paper, we analyzed in detail the light curves and eclipse timings of the W UMa system KIC 9532219, based on the high 
quality {\it Kepler} data from Quarter 3 through Quarter 17. The results from these analyses can be summarized as follows:

\begin{enumerate}
\item KIC 9532219 exhibits significant changes in the eclipse depths and timings. The {\it Kepler} light curves are satisfactorily
fitted by using third-body parameters and by adopting a magnetic cool spot on the less massive primary star. The result indicates 
that the eclipsing pair of the system is a marginal contact binary in which both components fill 99.9 \% of their inner Roche lobes.

\item Our syntheses for the 312 light curves, combined at intervals of about 4 d, indicate that the change of $l_{1}$ is nearly 
180$^\circ$ out of phase with that of $l_{3}$ and the two parameters have varied in patterns very similar to the depth changes 
of the eclipses. The variable asymmetries of light maxima, the so-called variable O'Connell effect, result from the changes in 
the longitude and temperature factor of the cool spot with time, which do not significantly contribute to the variations in 
eclipse depth. The observed eclipse depths may be severely affected by the presence of the nearby star KIC 9532228 and 
the calculated luminosities seem to be modulated at the {\it Kepler} orbital period of 372.5 d. These indicate that eclipse depth 
variation could be produced by changing amounts of contamination due to the nearby star associated with the orbital motion of 
the {\it Kepler} spacecraft.

\item The eclipse timings of KIC 9532219 have varied due to a combination of a downward parabola and a sinusoid with 
a period of 1196 d and a semi-amplitude of 69.6 s, rather than a period modulation suggested by Conroy et al. (2014). 
The periodic component is interpreted as the LTT effect due to a circumbinary companion. The mass function of this object is 
$f(M_{3})$=0.000255 $M_\odot$, and the mass is calculated to be $M_{3} \sin i_{3}$=0.0892 $M_\odot$. This value is close to 
the theoretical limit of $\sim$0.072 M$_\odot$ for a brown dwarf star. The third body contributes very little to the total light 
of this system.

\item Although the estimated masses and radii of the system contain large uncertainties, it might be difficult that 
the parabolic variation with a period decrease rate of $-$5.27$\times$10$^{-7}$ d yr$^{-1}$ is explained by either the secondary 
to primary mass transfer or AML due to a magnetic stellar wind braking. We suggest that it may only be the observed part of 
a second LTT orbit caused by the presence of another companion in a wider orbit. Then, the hypothetical fourth component could 
be the main source of the third light ({\it $l_{3}$}$\sim$76\%) detected in our light-curve synthesis, together with the fainter 
star KIC9532228 with a separation of 4 arcsec.
\end{enumerate}

Assuming that our interpretation of the light and timing variations of KIC 9532219 is correct, KIC 9532219 would be 
a quadruple system, which consists of an eclipsing binary and two circumbinary companions. The third and fourth components 
might provide a significant clue to the formation and evolution of the ultrashort period binary and could be the main source 
of the third light detected in our syntheses. The presence of the circumbinary objects and the consequent contribution of 
the third light to the system luminosity would make KIC 9532219 an interesting target for dynamical evolutionary studies of 
multiple systems. A large number of future accurate mid-eclipse times are needed to identify the presence of 
the supposed fourth component. Because this system includes a faint binary with a very short orbital period of about 4.8 h, 
10-m class telescopes, such as Keck and GTC, will help to measure its radial velocities.

\acknowledgments{ }
We appreciate the careful reading and valuable comments of the anonymous referee. This paper includes data collected by 
the {\it Kepler} mission. {\it Kepler} was selected as the 10th mission of the Discovery Program. Funding for the {\it Kepler} 
mission is provided by the NASA Science Mission directorate. We have used the Simbad database maintained at CDS, Strasbourg, 
France. This work was supported by the KASI (Korea Astronomy and Space Science Institute) grant 2016-1-832-01.

\clearpage
\begin{figure}
\includegraphics[]{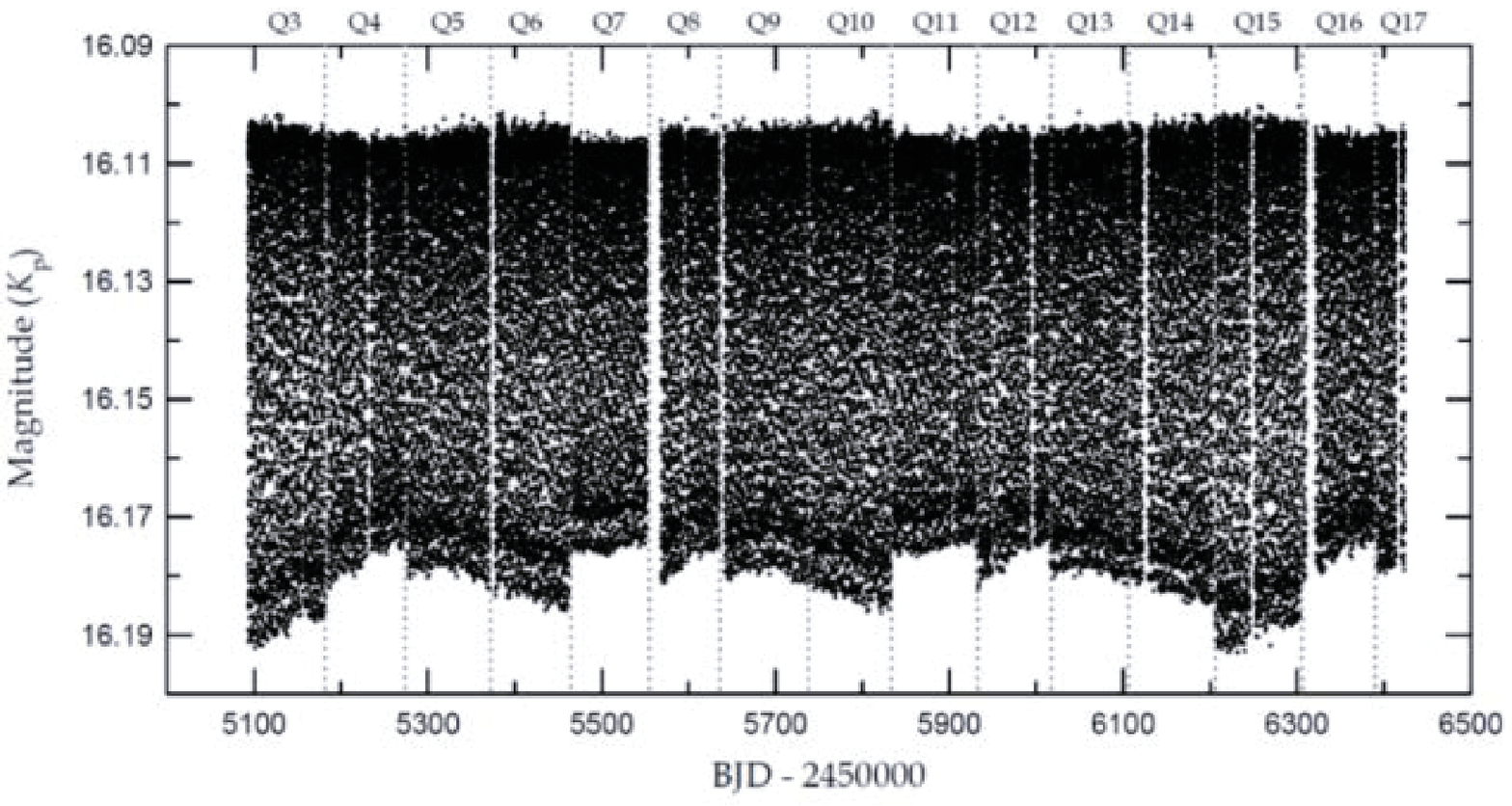}
\caption{Observed {\it Kepler} light curve of KIC 9532219. It can be seen that the depth of the eclipse has changed with time 
during $\sim$ 3.6 yr. The vertical dashed lines represent the ending times of each Quarter. }
\label{Fig1}
\end{figure}

\begin{figure}
\includegraphics[]{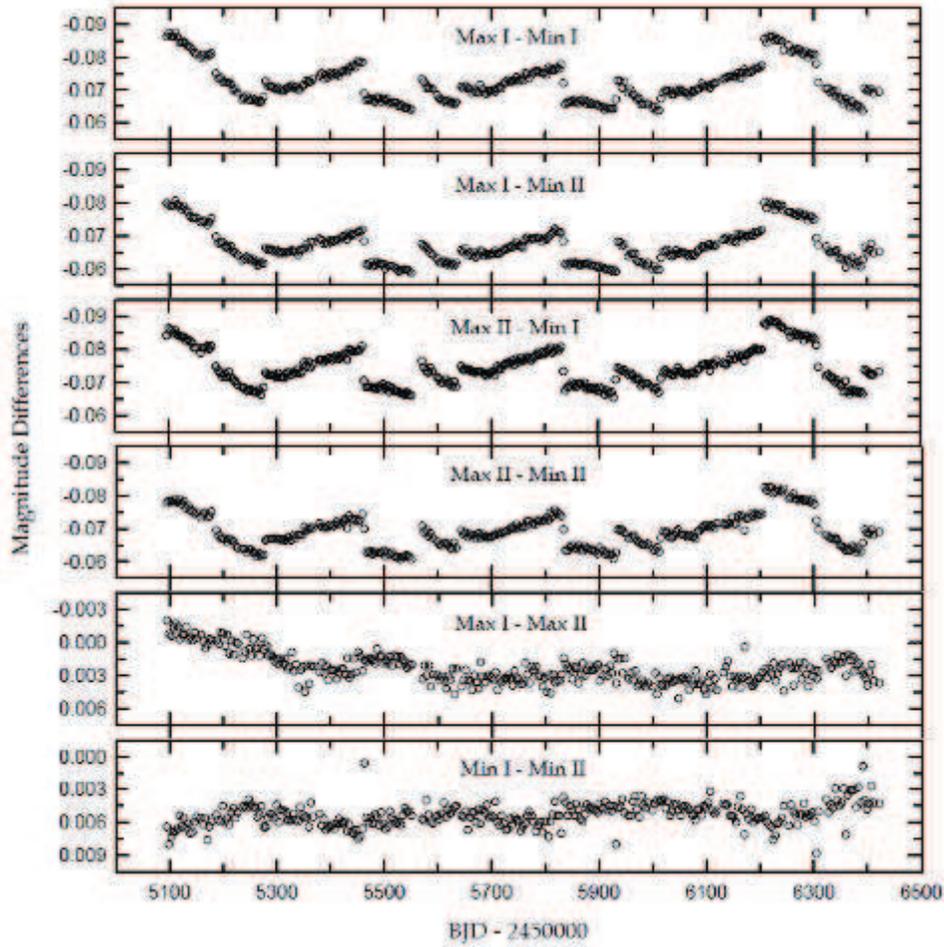}
\caption{Variations of the magnitude differences among four characteristic phases: Min I, Max I, Min II, and Max II. }
\label{Fig2}
\end{figure}

\begin{figure}
\includegraphics[]{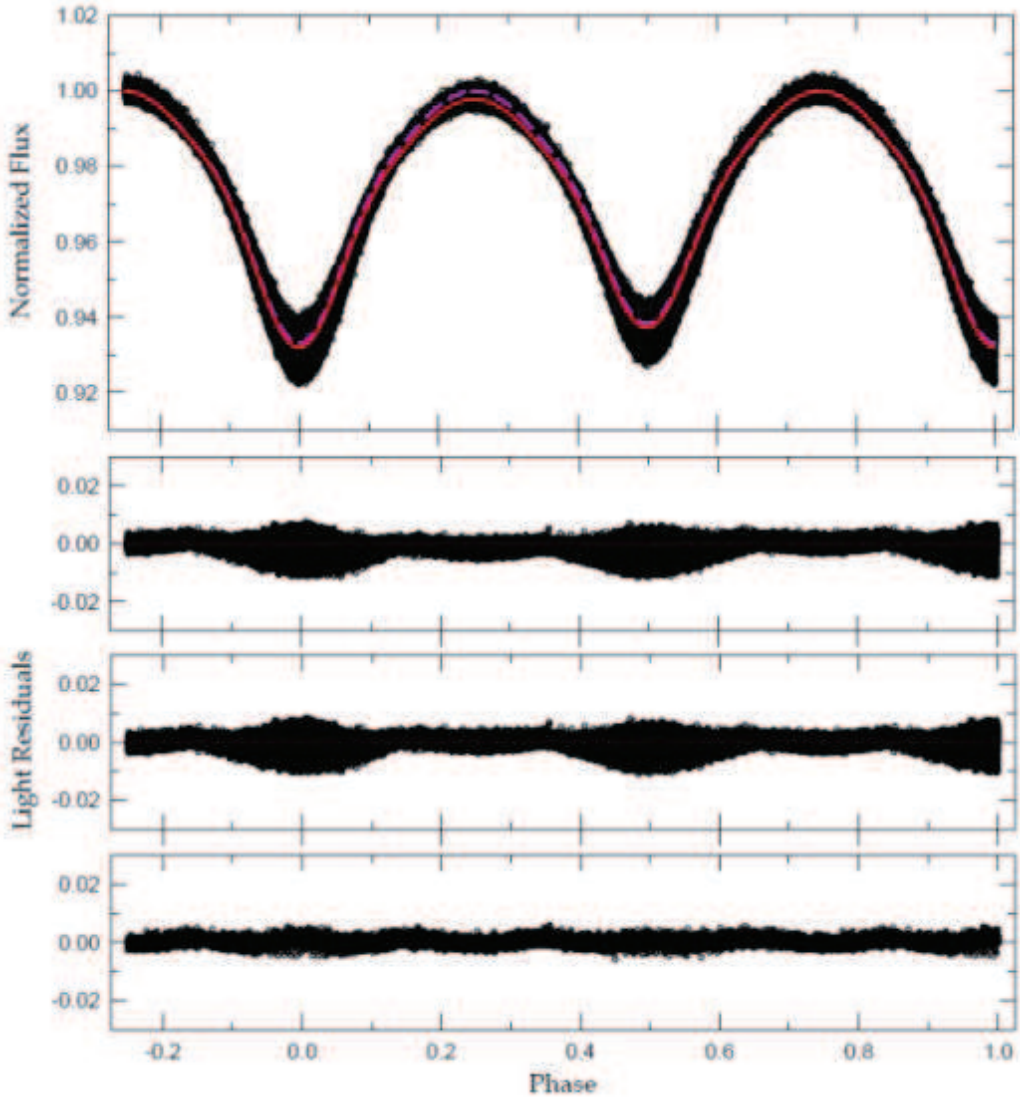}
\caption{Phased light curve of KIC 9532219 with the fitted models. The circles are individual measures from the {\it Kepler} 
satellite and the dashed and solid curves are computed without and with a cool spot on the primary star, respectively, 
in the simultaneous analyses of all observations. The light residuals corresponding to the unspotted and cool-spot models are 
plotted in the second and third panels, respectively. The bottom panel represents the residuals from each spot solution for 
the 312 datasets. }
\label{Fig3}
\end{figure}

\begin{figure}
\includegraphics[]{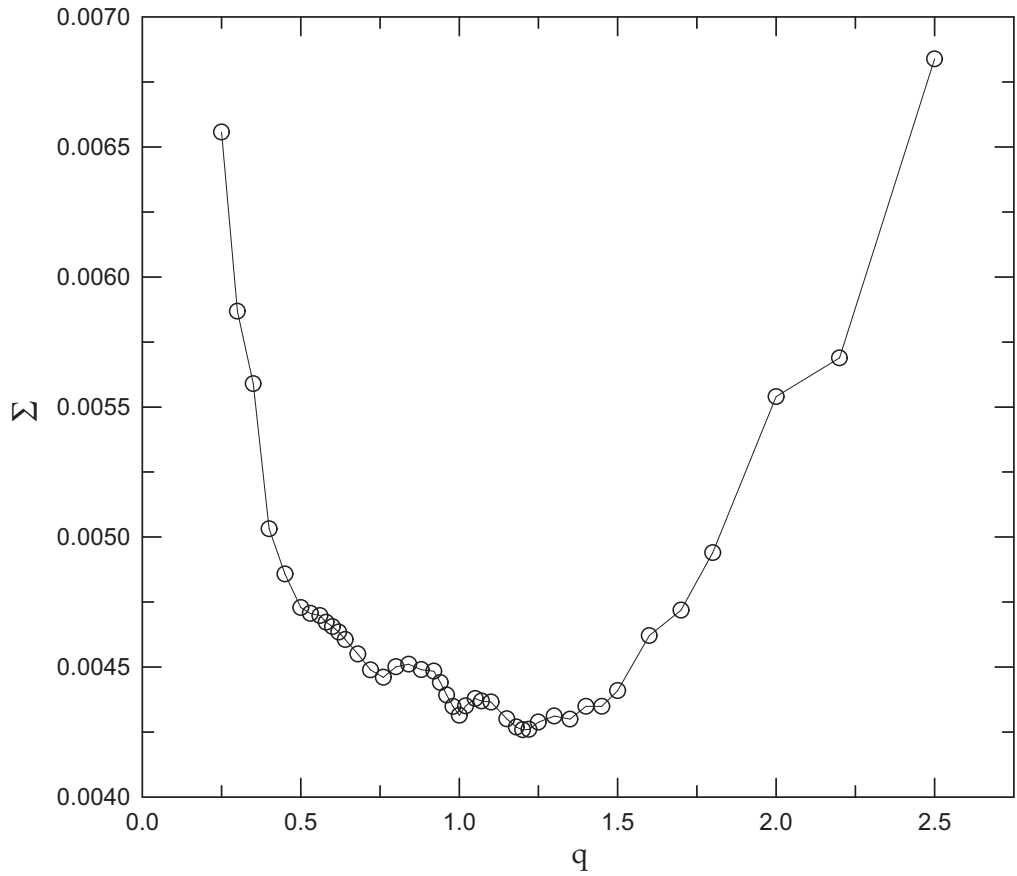}
\caption{Behavior of $\Sigma$ (the weighted sum of the residuals squared) of KIC 9532219 as a function of mass ratio $q$, showing 
a minimum value at $q$=1.20. }
\label{Fig4}
\end{figure}

\begin{figure}
\includegraphics[]{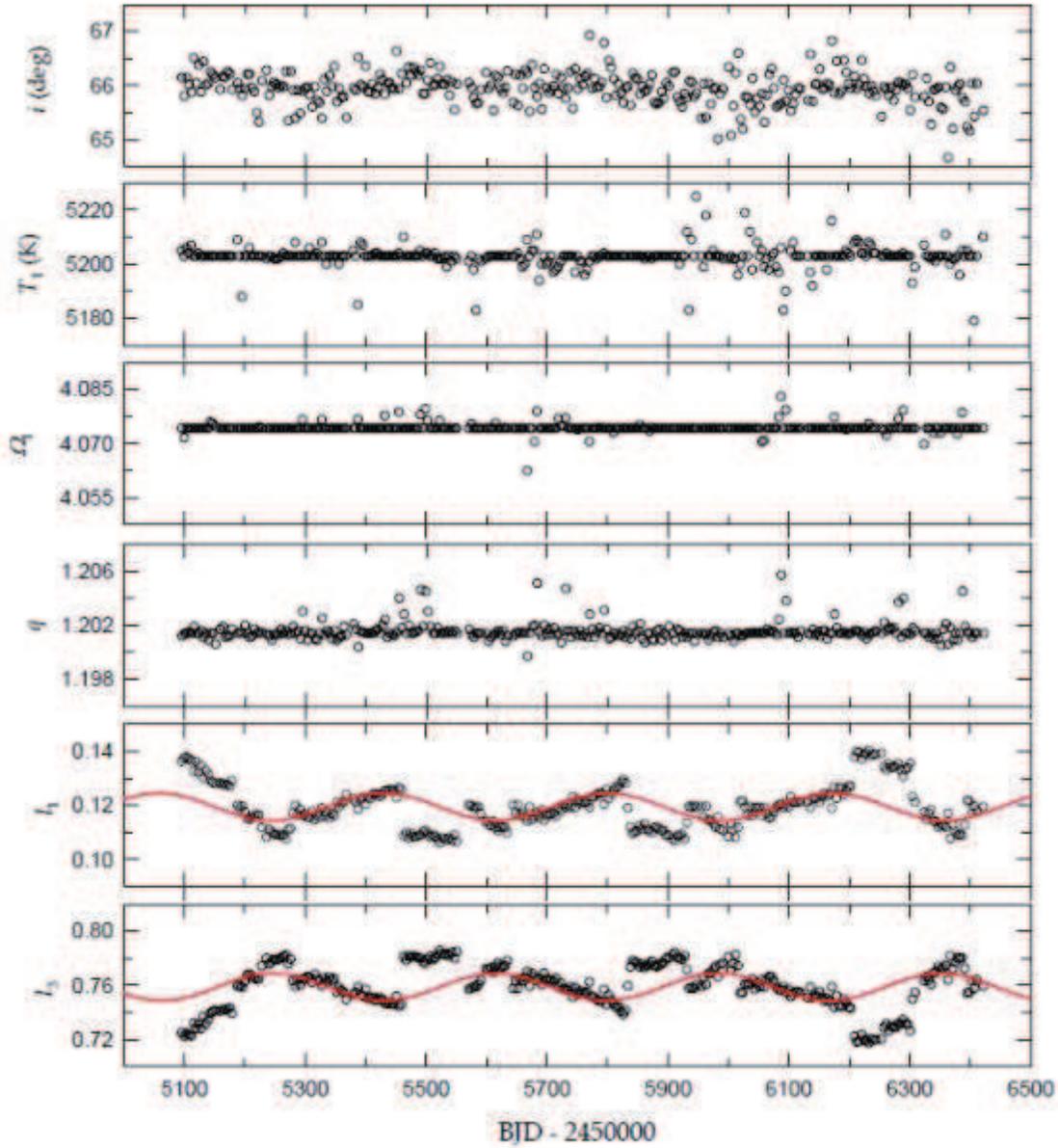}
\caption{Variations in the binary parameters of KIC 9532219 calculated from the 312 light curves. The solid curves in the fifth 
to sixth panels represent the results obtained by fitting a sine wave to the $l_1$ and $l_3$ values, respectively. }
\label{Fig5}
\end{figure}

\begin{figure}
\includegraphics[]{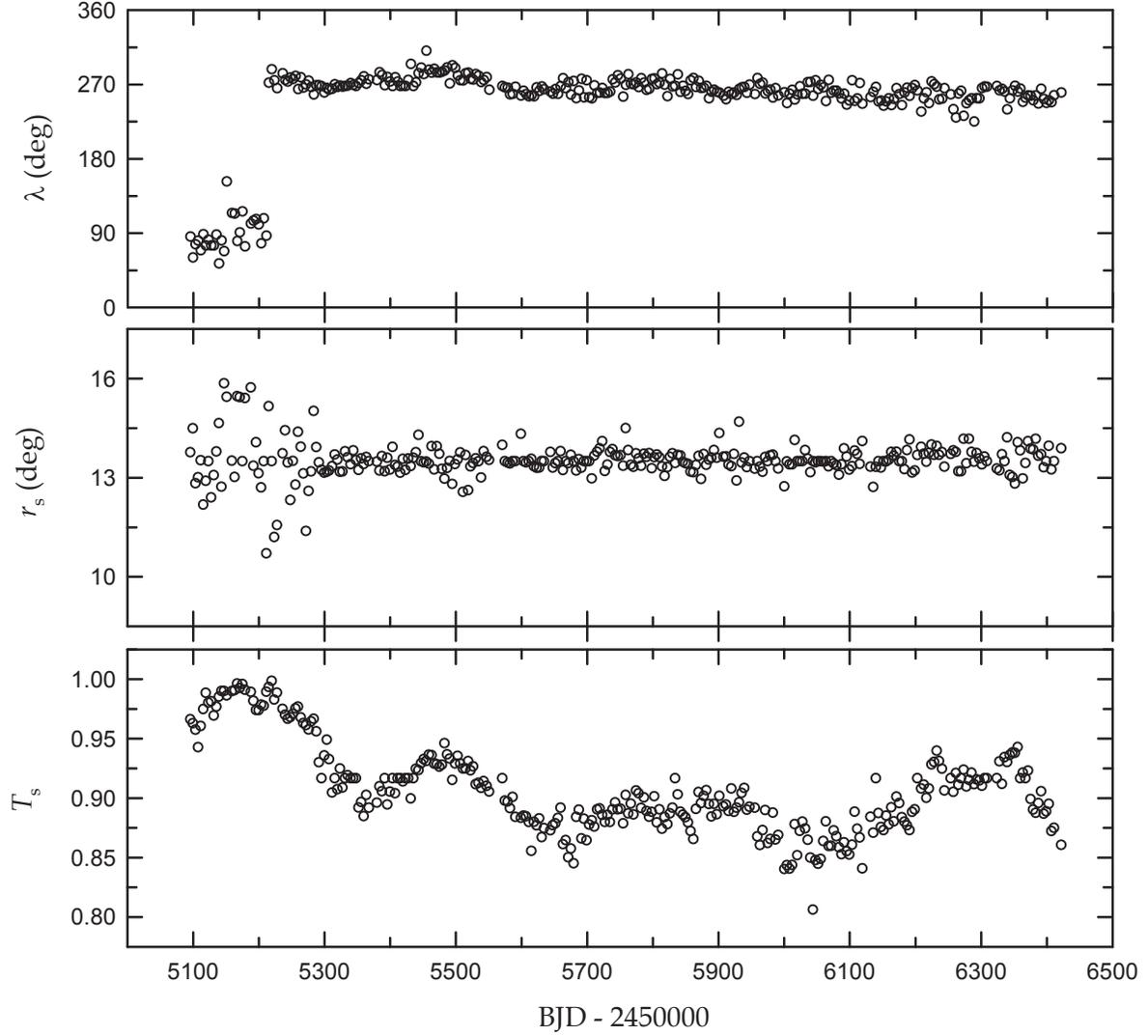}
\caption{Variations in the spot parameters of KIC 9532219 calculated from the 312 light curves. }
\label{Fig5}
\end{figure}

\begin{figure}
\includegraphics[]{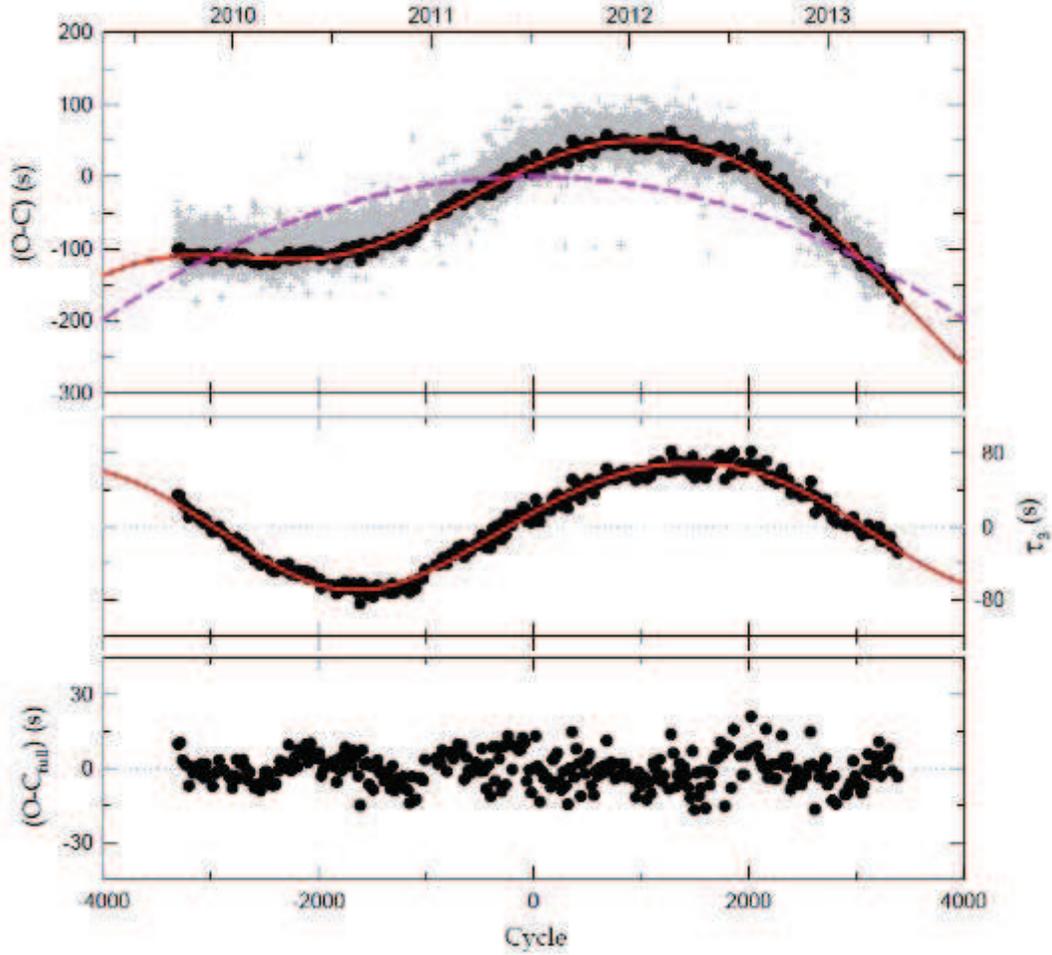}
\caption{$O$--$C$ diagram of KIC 9532219 with respect to the linear terms in Table 5. The plus symbols and open circles are 
the measures of Conroy et al. (2014) and ours, respectively. In the top panel, the solid and dashed curves represent 
the full contribution and just the quadratic term ($A$), respectively. The middle panel displays the LTT orbit ($\tau_{3}$) 
and the bottom panel the residuals from the complete ephemeris. }
\label{Fig7}
\end{figure}

\clearpage
\begin{deluxetable}{lcccc}
\tablewidth{0pt}
\tablecaption{Light levels of KIC 9532219 at four characteristic phases.}
\tablehead{
\colhead{Mean Time}  & \colhead{Min I}    & \colhead{Max I}    & \colhead{Min II}   & \colhead{Max II}    \\
\colhead{(BJD)}      & \colhead{(mag)}    & \colhead{(mag)}    & \colhead{(mag)}    & \colhead{(mag)}     
}
\startdata
2,455,095.21711      &  16.190$\pm$0.001  &  16.104$\pm$0.001  &  16.184$\pm$0.001  &  16.106$\pm$0.001   \\
2,455,099.18112      &  16.191$\pm$0.002  &  16.104$\pm$0.002  &  16.183$\pm$0.002  &  16.104$\pm$0.001   \\
2,455,103.14512      &  16.191$\pm$0.002  &  16.104$\pm$0.002  &  16.183$\pm$0.002  &  16.105$\pm$0.001   \\
2,455,107.10912      &  16.191$\pm$0.002  &  16.105$\pm$0.002  &  16.184$\pm$0.002  &  16.106$\pm$0.002   \\
2,455,111.07310      &  16.191$\pm$0.002  &  16.103$\pm$0.002  &  16.184$\pm$0.003  &  16.105$\pm$0.002   \\
2,455,115.03709      &  16.190$\pm$0.002  &  16.104$\pm$0.002  &  16.183$\pm$0.002  &  16.105$\pm$0.002   \\
2,455,119.00106      &  16.188$\pm$0.002  &  16.104$\pm$0.002  &  16.183$\pm$0.002  &  16.104$\pm$0.002   \\
2,455,122.96503      &  16.190$\pm$0.004  &  16.105$\pm$0.003  &  16.184$\pm$0.003  &  16.106$\pm$0.002   \\
2,455,126.92900      &  16.189$\pm$0.003  &  16.104$\pm$0.003  &  16.183$\pm$0.003  &  16.105$\pm$0.003   \\
2,455,130.89296      &  16.188$\pm$0.002  &  16.104$\pm$0.003  &  16.181$\pm$0.003  &  16.105$\pm$0.004   \\
\enddata
\tablecomments{This table is available in its entirety in machine-readable and Virtual Observatory (VO) forms in the online journal. 
A portion is shown here for guidance regarding its form and content.}
\end{deluxetable}

\begin{deluxetable}{lccccc}
\tabletypesize{\small}  
\tablewidth{0pt} 
\tablecaption{Binary parameters of KIC 9532219 obtained by fitting simultaneously all {\it Kepler} data }
\tablehead{
\colhead{Parameter}                      & \multicolumn{2}{c}{Unspotted Model}             && \multicolumn{2}{c}{Cool-Spot Model}               \\ [1.0mm] \cline{2-3} \cline{5-6} \\[-2.0ex]
                                         & \colhead{Primary} & \colhead{Secondary}         && \colhead{Primary} & \colhead{Secondary} 
}
\startdata                                                                                                                                      
$T_0$ (BJD)                              & \multicolumn{2}{c}{2,455,750.57728(4)}          && \multicolumn{2}{c}{2,455,750.57728(3)}            \\
$P$ (d)                                  & \multicolumn{2}{c}{0.19815494(5)}               && \multicolumn{2}{c}{0.19815494(5)}                 \\
d$P$/d$t$                                & \multicolumn{2}{c}{$-$1.4(1)$\times$10$^{-9}$}  && \multicolumn{2}{c}{$-$1.4(1)$\times$10$^{-9}$}    \\
$q$                                      & \multicolumn{2}{c}{1.201(2)}                    && \multicolumn{2}{c}{1.201(1)}                      \\
$i$ (deg)                                & \multicolumn{2}{c}{66.0(3)}                     && \multicolumn{2}{c}{66.0(3)}                      \\
$T$ (K)                                  & 5,203(23)         & 5,031                       && 5,203(20)         & 5,031                         \\
$\Omega$                                 & 4.0743(6)         & 4.0743                      && 4.0742(5)         & 4.0743                        \\
$X$, $Y$                                 & 0.644, 0.175      & 0.642, 0.166                && 0.644, 0.175      & 0.642, 0.166                  \\
$x$, $y$                                 & 0.704, 0.198      & 0.712, 0.176                && 0.704, 0.198      & 0.712, 0.176                  \\
$l$/($l_{1}$+$l_{2}$+$l_{3}$)            & 0.1181(2)         & 0.1203                      && 0.1195(2)         & 0.1216                        \\
$l_{3}$$\rm ^a$                          & \multicolumn{2}{c}{0.7616(5)}                   && \multicolumn{2}{c}{0.7589(4)}                     \\
$r$ (pole)                               & 0.3404(1)         & 0.3710(1)                   && 0.3405(1)         & 0.3711(1)                     \\
$r$ (side)                               & 0.3567(1)         & 0.3905(1)                   && 0.3568(1)         & 0.3906(1)                     \\
$r$ (back)                               & 0.3880(1)         & 0.4205(1)                   && 0.3881(1)         & 0.4207(1)                     \\
$r$ (volume)$\rm ^b$                     & 0.3631            & 0.3954                      && 0.3632            & 0.3956                        \\ [1.0mm]
\multicolumn{6}{l}{Third-body parameters:}                                                                                                      \\        
$a^{\prime}$($R_\odot$)                  & \multicolumn{2}{c}{560(25)}                     && \multicolumn{2}{c}{560(25)}                       \\        
$i^{\prime}$ (deg)                       & \multicolumn{2}{c}{66.0}                        && \multicolumn{2}{c}{66.0}                          \\        
$e^{\prime}$                             & \multicolumn{2}{c}{0.13(5)}                     && \multicolumn{2}{c}{0.13(4)}                       \\        
$\omega^{\prime}$  (deg)                 & \multicolumn{2}{c}{138(22)}                     && \multicolumn{2}{c}{134(21)}                       \\        
$P^{\prime}$ (d)                         & \multicolumn{2}{c}{1190(79)}                    && \multicolumn{2}{c}{1190(76)}                      \\        
$T_{\rm c}^{\prime}$ (BJD)               & \multicolumn{2}{c}{2,454,870(85)}               && \multicolumn{2}{c}{2,454,866(82)}                 \\ [1.0mm]
\multicolumn{6}{l}{Spot parameters:}                                                                                                            \\        
Colatitude $\theta$ (deg)                & \dots             & \dots                       && 86.0(7)           & \dots                         \\        
Longitude $\lambda$ (deg)                & \dots             & \dots                       && 268.2(3)          & \dots                         \\        
Radius $r_{\rm s}$ (deg)                 & \dots             & \dots                       && 13.5(2)           & \dots                         \\        
Temperature Factor $T_{\rm s}$           & \dots             & \dots                       && 0.917(2)          & \dots                         \\
$\Sigma W(O-C)^2$                        & \multicolumn{2}{c}{0.00233}                     && \multicolumn{2}{c}{0.00221}                       \\ 
\enddata
\tablenotetext{a}{Value at 0.75 phase.}
\tablenotetext{b}{Mean volume radius.}
\end{deluxetable}

\begin{deluxetable}{lrcccccccc}
\tabletypesize{\small}
\tablewidth{0pt}
\tablecaption{Light-curve solutions for the 312 datasets combined at intervals of 20 orbital periods.}
\tablehead{
\colhead{Epoch}    & \colhead{$\lambda$}  & \colhead{$r_{\rm s}$}   & \colhead{$T_{\rm s}$}   & \colhead{$i$}    & \colhead{$T_1$}  & \colhead{$\Omega_1$}  & \colhead{$q$}  & \colhead{$l_1$}  & \colhead{$l_3$}   \\
\colhead{(BJD)}    & \colhead{(deg)}      & \colhead{(deg)}         &                         & \colhead{(deg)}  & \colhead{(K)}    &                       &                &                  &                     
}
\startdata
2,455,095.277651   &  85.71               & 13.77                   & 0.966                   & 66.15            & 5205             & 4.0742                & 1.2012         & 0.1363           & 0.7251            \\
2,455,099.240760   &  60.67               & 14.50                   & 0.963                   & 65.81            & 5203             & 4.0716                & 1.2014         & 0.1375           & 0.7226            \\
2,455,103.203780   &  76.34               & 12.83                   & 0.958                   & 66.15            & 5206             & 4.0742                & 1.2014         & 0.1379           & 0.7247            \\
2,455,107.166866   &  80.99               & 13.01                   & 0.943                   & 66.02            & 5204             & 4.0742                & 1.2015         & 0.1372           & 0.7233            \\
2,455,111.129957   &  69.29               & 13.53                   & 0.961                   & 65.87            & 5207             & 4.0742                & 1.2014         & 0.1366           & 0.7228            \\
2,455,115.092961   &  88.67               & 12.18                   & 0.975                   & 66.52            & 5203             & 4.0742                & 1.2019         & 0.1326           & 0.7323            \\
2,455,119.056151   &  75.05               & 12.90                   & 0.989                   & 66.04            & 5203             & 4.0742                & 1.2014         & 0.1363           & 0.7284            \\
2,455,123.019220   &  82.01               & 13.50                   & 0.980                   & 66.38            & 5204             & 4.0742                & 1.2014         & 0.1323           & 0.7313            \\
2,455,126.982340   &  75.05               & 12.40                   & 0.981                   & 65.88            & 5203             & 4.0742                & 1.2016         & 0.1348           & 0.7280            \\
2,455,130.945450   &  75.05               & 13.08                   & 0.969                   & 66.45            & 5203             & 4.0742                & 1.2011         & 0.1302           & 0.7372            \\
\enddata
\tablecomments{This table is available in its entirety in machine-readable and Virtual Observatory (VO) forms in the online journal. 
A portion is shown here for guidance regarding its form and content.}
\end{deluxetable}

\begin{deluxetable}{lcc}
\tablewidth{0pt} 
\tablecaption{Absolute dimensions for KIC 9532219.}
\tablehead{
\colhead{Parameter}              & \colhead{Primary} & \colhead{Secondary}
}
\startdata
$M$($M_\odot$)                   & 0.72              &  0.86            \\
$R$($R_\odot$)                   & 0.60              &  0.66            \\
$\log$ $g$ (cgs)                 & 4.73              &  4.74            \\
$L$($L_\odot$)                   & 0.24              &  0.25            \\
$M_{\rm bol}$ (mag)              & 6.28              &  6.24            \\
BC (mag)                         & $-$0.23           &  $-$0.29         \\
$M_{\rm V}$ (mag)                & 6.51              &  6.53            \\
\enddata
\end{deluxetable}

\begin{deluxetable}{lccccc}
\tablewidth{0pt}
\tablecaption{Parameters for the quadratic {\it plus} LTT ephemeris of KIC 9532219}
\tablehead{
\colhead{Parameter}     & \colhead{Value}              & \colhead{Unit}   \\
}                                                                                                                                         
\startdata                                                                                                                                
$T_0$                   & 2,455,750.5771475(42)        & BJD              \\
$P$                     & 0.1981549231(22)             & d                \\
$A$                     & $-1.4286(85)\times 10^{-9}$  & d                \\
$a_{\rm b}\sin i_{3}$   & 0.1398(11)                   & au               \\
$e_{\rm b}$             & 0.1504(85)                   &                  \\
$\omega_{\rm b}$        & 247.83(41)                   & deg              \\
$n_{\rm b}$             & 0.30110(73)                  & deg d$^{-1}$     \\
$T_{\rm b}$             & 2,455,361.1(1.3)             & BJD              \\
$P_{\rm b}$             & 1195.6(2.9)                  & d                \\
$K_{\rm b}$             & 69.64(54)                    & s                \\
$f(M_{3})$              & 0.0002548(21)                & M$_\odot$        \\ 
$M_{3} \sin i_{3}$      & 0.08921(35)                  & M$_\odot$        \\
$a_{3} \sin i_{3}$      & 2.4755(49)                   & au               \\
$e_3$                   & 0.1504(85)                   &                  \\
$\omega_{3}$            & 67.83(41)                    & deg              \\
$P_3$                   & 1195.6(2.9)                  & d                \\
$dP$/$dt$               & $+5.267(31)\times 10^{-7}$   & d yr$^{-1}$      \\[0.5mm]
rms scatter             & 6.56                         & s                \\
$\chi^2 _{\rm red}$     & 1.013                        &                  \\
\enddata
\end{deluxetable}

\end{document}